# Atomic Batteries: Energy from Radioactivity

Suhas Kumar

*Department of Electrical Engineering, Stanford University, Stanford, CA 94305, USA, email:* su1@alumni.stanford.edu

*17 Nov, 2015*

**With alternate, sustainable, natural sources of energy being sought after, there is new interest in energy from radioactivity, including natural and waste radioactive materials. A study of various atomic batteries is presented with perspectives of development and comparisons of performance parameters and cost. We discuss radioisotope thermal generators, indirect conversion batteries, direct conversion batteries, and direct charge batteries. We qualitatively describe their principles of operation and their applications. We project possible market trends through our comparative cost analysis. We also explore a future direction for certain atomic batteries by using nanomaterials to improve their performance.**

Atomic batteries, nuclear batteries or radioisotope generators are devices that use energy from radioactive decay to generate electricity. Similar to nuclear reactors, they generate electricity from atomic energy, but differ in that they do not use chain reactions and instead use continual radioactive emissions to generate electricity. One of the earliest efforts to make such a battery was in 1913.[1] The two primary types of radioactive decay, alpha decay and beta decay, can be visualized as shown in Figure 1.

There have been several motivations for people to have pursued radioisotope batteries for about a century now. An important factor is the longevity of these systems, where the life of the battery is a strong function of the half-life of the material used, which can easily be in the order of many decades. These batteries have high energy density, up to five orders higher than chemical batteries.[2] These systems can function over wide ranges of environmental conditions of temperature, pressure, under water or in space. Since radioisotope decay is sustained by the material itself, there is no need for refueling or recharging. The downside of these batteries is that their power density is lower or comparable with chemical batteries. They also have low conversion efficiencies –10% efficiency would be considered a great atomic battery. Another motivation is that these can be made from the waste of nuclear fission. Using radioactive materials also poses issues with regulations of usage and disposal. These batteries enable compact and high energy capacity power generators for applications ranging from implantable cardiac pacemakers[3] to space stations.[4] Currently, radioisotope power generators are being developed[5] for realizing safe, compact, high energy capacity, and long lifetime batteries for remote wireless sensor microsystems in applications ranging from environmental health monitoring to structural health monitoring.[6] These are also employed in a variety of industrial applications including electron capture devices for gas chromatography.[7]

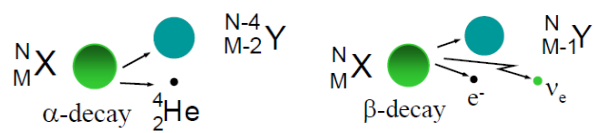

**Figure 1:** Visualization of (left) alpha decay and (right) beta decay.

Let us do a calculation to check the reason for such applications. Consider 1000 kg of U-235, which is used in nuclear plants, that has a half life of $4.5 \times 10^9$ years; the energy released in each alpha particle is 4.27MeV.[8] The decay constant, $k=\ln2/T_{1/2}$ ~ $4.9 \times 10^{-18}$. The number of nuclei, N, is 1000kg/(235*mass of proton) ~ $2.6 \times 10^{27}$. So the activity is $k*N$ ~ $1.3 \times 10^{10}$ s$^{-1}$. Now, power is activity*energy per particle, which is about 0.0085 W. If we convert all this energy to electricity, we have barely enough to power an LED! If we replace

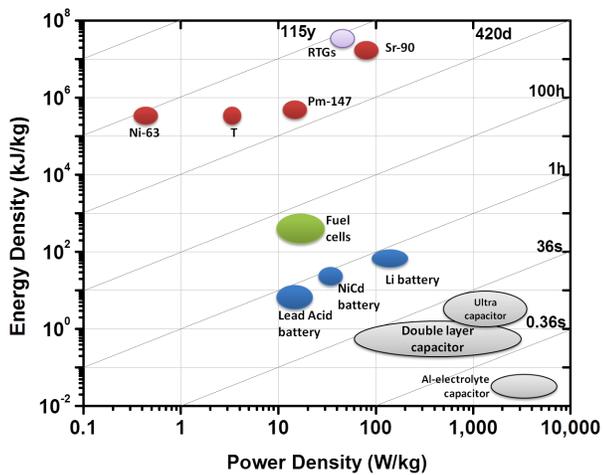

**Figure 2:** Ragone plot showing capacitors (gray), chemical batteries (blue), fuel cell (green), atomic batteries of various radioisotopes (red) and RTGs (purple). The sloped lines are constant-time lines. Data collected from Refs. 2, 10.

U-235 with Cs-137 with a half life of about 30 years, it would yield 1 MW of power, which is sizable, yet not even close enough to run a power plant. This hints at the domains of operation: long lasting power supply, low power, high energy. Electrical technology being mature, we consider harnessing and/or storage in terms of electricity. We construct a Ragone plot for these batteries, shown in Figure 2. A Ragone plot is one which generally plots the energy density and power density on either of the axes.[9] Sloped lines in the plot are constant-time lines.

Conversion of radioisotope decay to electricity can be broadly classified into two types: thermal conversion (where the thermal power of ionizing radiation is used) and non-thermal conversion (where the output does not depend on the thermal power of the source). Figure 3 summarizes the methods of conversion of radioactive decay to electricity. The choice of materials for these batteries depends on the power and energy density considerations, which can be inferred from the Ragone plot for the individual materials. Table 1 shows some of these numbers.

**RADIOISOTOPE THERMAL GENERATORS (RTGs)**

Thermal converters (radioisotope thermoelectric generators - RTGs) use the thermal energy of the radioisotope decay to generate electricity. Methods to accomplish this include heating up a thermocouple, producing infrared radiation from hot metals to power 'solar cells', using the Stirling Engine and many more.

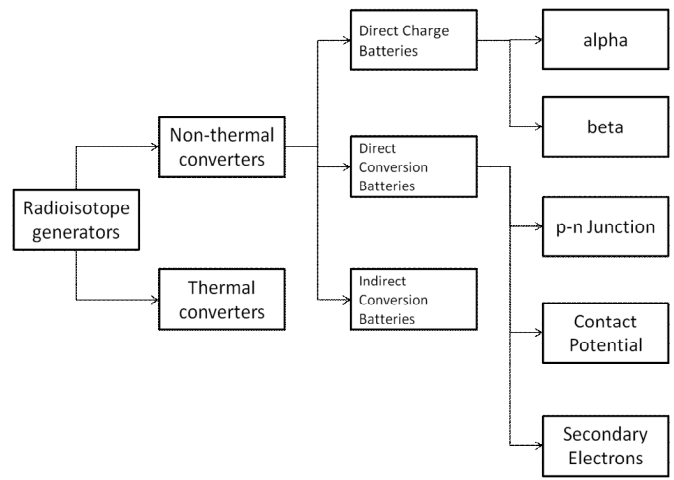

**Figure 3:** Types and classification of relevant radioisotope generators.[2, 4]

These are effective starting at several hundred milliwatt of electrical power. A large amount of radioactive material is necessary to create a sufficient thermal gradient for an effective RTG; at least a gram of the alpha or beta radioactive isotopes, with emitted particle energy of several hundreds or thousands keV (usually Pu-238 and Sr-90)[4], are used. The efficiency of energy conversion for RTGs can reach 8-10%.[11] Some modern RTG thermo-photovoltaic cells can reach conversion efficiency up to 20% and theoretical efficiency reaches 30%.[12] Prototypes for the new generation of RTG, Stirling Radioisotope Generator, demonstrated an average efficiency of 23%.[2] The large amount of radioactive isotopes in RTGs restrict their applications because of high radiation dangers. NASA has been using the thermocouple based RTG for the past 30 years or so. Figure 4 shows a photograph of an RTG that NASA's Apollo 14 mission carried to the Moon. Since it proved too heavy, expensive and inefficient, they are replacing it with a far better version of an RTG – the RTG based Stirling Engine.[13] Patented in 1816 by Robert Stirling, the Stirling engine consists of two chambers or cylinders, one cold and one hot, contains a "working fluid" (commonly air, helium or hydrogen) with a regenerator or heat exchanger between the two. Differences in temperature and pressure between the two cylinders cause the working fluid to expand and contract, passing back and forth through the exchanger and moving a piston. The process hence converts thermal energy (in NASA's case, supplied by radioactivity) to mechanical energy.

|  | Isotope | | | | |
| Parameter | Tritium | Pm-147 | Ni-63 | Sr-90 | Pu-238 |
|---|---|---|---|---|---|
| Half-life of isotope, $T_{1/2}$, yr | 12.32 | 2.62 | 100.1 | 28.9 | 87.7 |
| Chemical compound of isotope | $Ti_3H_2$, $Sc_3H_2$ | $^{147}Pm_2O_3$ | $^{63}Ni$ | $^{90}Sr(NO_3)_2$ | $^{238}PuO_2$ |
| Specific activity of the compound, $A_{sp}$, Ci/g | 1100 | 800 | 57 | 116 | 15 |
| Specific power of isotope, $P_0$, μW/Ci | 34 | 367 | 103 | 6700 | 32000 |

**Table 1:** Parameter values for materials popularly considered for atomic batteries.[2, 4, 8]

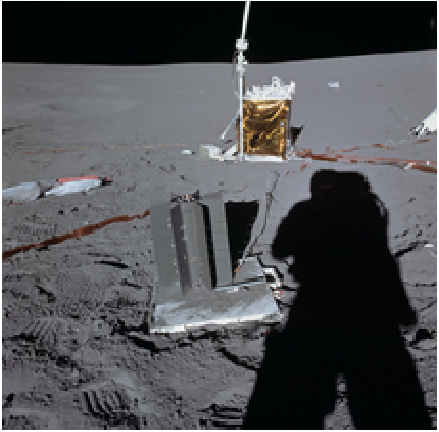

**Figure 4:** A photograph of the RTG that NASA's Apollo 14 mission carried to the Moon. The RTG is the gray colored device with cooling fins. Source: Wikimedia Commons.

### INDIRECT CONVERSION BATTERIES

Indirect conversion typically involves two steps of conversion. The radioactive decay consisting of either alpha or beta particles is impinged on some radio luminescent material like phosphor to produce ultra photons and then is collected using photodiodes or 'solar cells'. The intensity of radio luminescent light source, on the high end, can be about 20μW/cm², which would require the use of photodiodes suitable for low intensity light.[14b] Also, the spectral response of the radio luminescent light source must be matched with the response of the material of the collector that absorbs the light, so as to ensure highest possible efficiency. We use III-V materials like InAs, GaP, GaAs, AlGaAs, etc., which are most suitable for photodiode applications in the visible and ultraviolet range.[15] Optimization is also done on the structure of the radio luminescent material. Some of the reported structures are nano tubular or micro spherical structures filled with tritium,[14b] thin films and nano sized powders. It is possible to play with the radio luminescent material too. There are reports that have used CdSe based phosphor, tritium in micro or nanosized particles and aerogel phosphor composition saturated with tritium or a tritium containing organic luminophor.[2] Figure 5 shows a schematic of a generic indirect converter and a specific example of using tubular structures filled with tritium as radio luminescent material and a III-V material, AlGaAs, as the photodiode material. It might also help to guide the photons to a remotely located photodiode using waveguides so that the diode is protected from the radiations. Such a schematic is also shown in Figure 5. At best, when the spectra are matched and the conversion of the power is optimal, we can expect an overall efficiency of 2% at 3.5V open circuit voltage.[14a] Theoretically this efficiency can be 25%, but it has not been shown experimentally yet.[16] The battery life depends on the half life of the radio isotope and also on how fast the radio luminescent material degrades with high energy radiation, which many a times turns out to be faster than the rate at which the radioactivity falls.[2]

### DIRECT CONVERSION BATTERIES

Direct conversion uses the radioisotope decay to directly drive a device that converts these charged particles to electricity. Typical methods include the 'betavoltaic' effect, contact potential difference and secondary emission from an irradiated surface.

The betavoltaic effect refers to using the beta particles of radioisotope decay to generate electricity. The betavoltaic effect refers to using the beta particles of radioisotope decay to generate electricity. This is typically done using a semiconductor junction. Early efforts of doing this can be traced back to 1953.[17] The theory behind this principle is very similar to the theory of solar cells or photovoltaic cells. Instead of having incoming photons create free electrons, we have found in Refs. 15 and 19. Energetic beta particles (typically a few keV) hit the semiconductor to produce

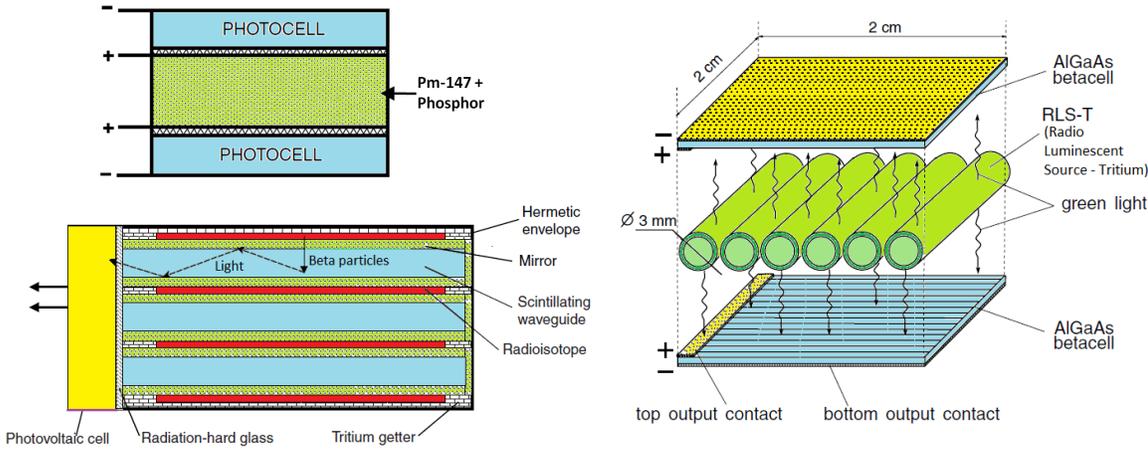

**Figure 5:** (top left) An indirect conversion battery with Pm-147 as the radioisotope and phosphorous as the source of radio-generated photons. (right) Radio-luminescent source is a bundle of tubular nanostructures filled with tritium, which produce visible light, green in this case, to drive photodiodes made of a III-V compound, AlGaAs. (bottom left) A setup that uses waveguides to guide all the collected photons to a diode. Modified and adapted from Ref. 14b.[14a, 14b]

hundreds of thousands of electron-hole pairs. If the device is well-designed, most of these free electrons can be swept across the electric field that's built in the p-n diode to produce current that can be fed to a load. Figure 6 depicts this process. Inevitably, there are some losses too. Some of the excited electrons undergo recombination, which is accelerated by traps and impurities in the crystal, to produce a photon and/or a phonon. Some beta particles undergo inelastic scattering, the energy of which goes up to heat up the crystal. Engineering the material itself is a fundamental process to improve the efficiency. III-V materials like GaAs, InP, etc., are suited for this application owing to their favorable and tunable band structure.[14, 15, 19] Semiconductor conversion efficiency for a single cell can reach 30%; this can be improved by modifications like stacking many cells. The product of source and conversion efficiencies can reach 10%, but it is experimentally shown only up to 2%.[2] There have been very few efforts to use nanomaterials to improve performance of direct conversion batteries, as is done in solar cells.[14, 18, 32] We discuss this domain in a later section. The open circuit voltage can reach many volts. The performance of these depend more on the degradation of the material upon being battered by radiation, since this process, like in RTGs, is faster than the radioisotope decay itself.[19] Liquid material can be used instead of solid to avoid damage.[20] Nonetheless; betavoltaic batteries have shown promise as on-chip batteries which have very long lifetimes.

A contact potential difference battery uses beta particles to generate electron-hole pairs in a gas or solid, which are held at two opposite ends by metals of different work functions. Owing to the difference in the work functions, the electrons move towards the low work-function electrode. This battery cannot reach more than 1V in open circuit and a couple of nano amperes in current. The efficiency is less than 1%.[21, 14c] This is similar to the concept of producing secondary electrons from the surface of a metal or the bulk of a dielectric and collecting them to produce current. An interesting deviation from these techniques is to use the gamma radiation from radioisotopes to generate current. This could be the Compton scattered electrons from gamma radiation that are stored in a dielectric. The energy in

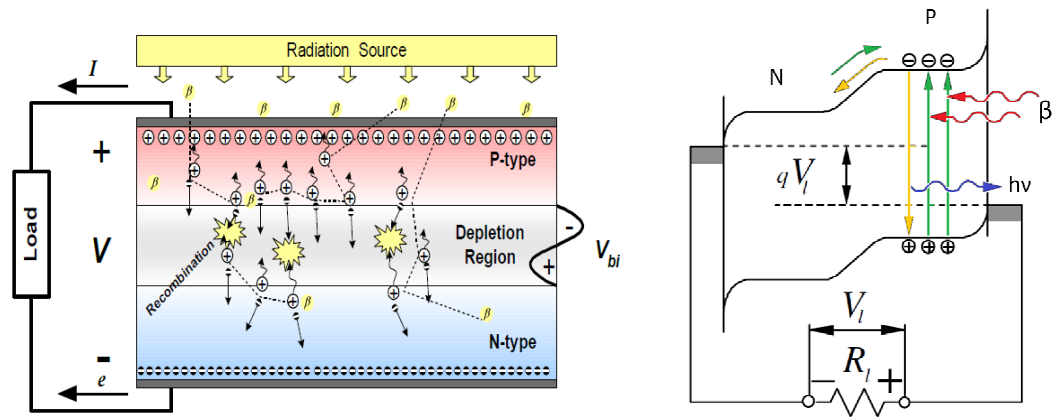

**Figure 6:** (left) Visualization of irradiation of beta particles generating electron hole pairs in a p-n junction, which are swept away by the junction to produce a current. (right) An equivalent band diagram picture of the process. Reproduced from Refs. 14b, 19.

this case is extremely high but the

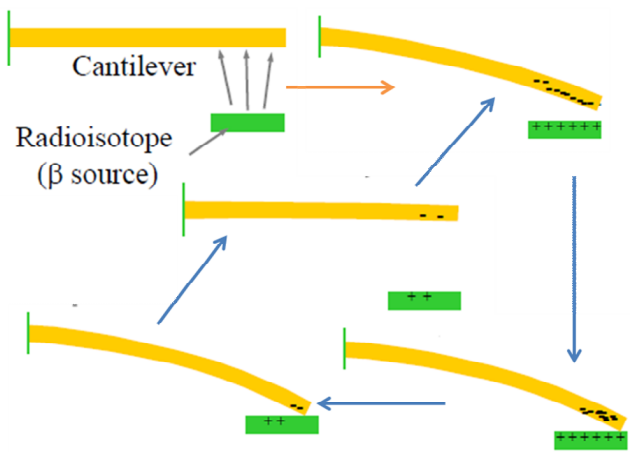

**Figure 7:** Working of the direct charge induction MEMS cantilever.

conversion is quite lossy. This kind of a cell could give up to 500,000V at 50mW of power, while the efficiency would be around 0.35%.[22] We do not often come across a gamma cell since it needs a heavy radiation shield to keep the gamma rays from getting out of the system.

### DIRECT CHARGE BATTERIES

Direct charge batteries rely on using the charge on beta or alpha particles to directly derive a current. The first atomic battery demonstrated by Moseley in 1913[1] was a direct charge battery. In simple terms it was a radioisotope held inside a metallic sphere without having electrical contact with the sphere. The radiating charged particles produce a voltage difference and a current when contacted. In Moseley's work, the battery could generate 150kV at about 0.01nA. The concept behind direct charge has come a long way since Moseley's experiment. One of the systems that has been getting a lot of attention lately is the MEMS (Micro electro mechanical system) – radioisotope coupled system.[23] As shown in Figure 7, a radioisotope produces charges that charge up a conductive beam. This builds up an increasing electric field that eventually pulls the beam down to contact the bottom surface, which allows the charges to be discharged. This produces a current. Electrostatic and dynamic analysis of this system is done in Ref. 24. An enhancement to this is to add a piezo electric material onto the cantilever. Every time it bends, it generates a piezo electric voltage. In addition to the discharge current, it can produce a time varying piezo electric voltage after the cantilever has sprung back and is oscillating towards its mean position at its natural frequency of resonance.[25] Figure 8 shows a schematic of the structure and the voltage profiles during actuation.

### COST ANALYSIS

We extract the costs involved for nuclear batteries and compare them to the corresponding values for other types of electrical storage. The costs that matter for production are the cost per kilogram ($/kg) of the material and the cost of setting up the processing facility to the corresponding materials. Additionally, costs that drive the battery market include cost per unit power ($/kW) and cost per unit energy ($/kJ) of the battery. We find costs of chemical batteries, fuel cells and capacitors by taking the ratio of cost to the weight ($/kg) of a packaged system in the market and using the energy and power density values we already know. Politics also influence the cost of nuclear batteries, since no government wants a free market for radioactive material and certainly not devices containing fertile material which can be bred to make a fission reaction, turning them into nuclear weapons; as a result, costs are found only in volatile documents. We found that batteries made of radioisotopes cost about $10^5$\$/kg,[26-30] irrespective of the material used. This constancy, disregarding the material properties, as noted in Table 1, for instance, may be due to the regulation on nuclear supplies. Our findings are shown in Figure 9, while a direct comparison of power density and energy density was shown in Figure 2. The $/kJ numbers reported in Figure 9 take into account the efficiency of the batteries as well.

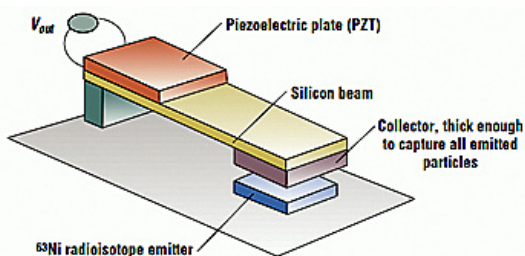
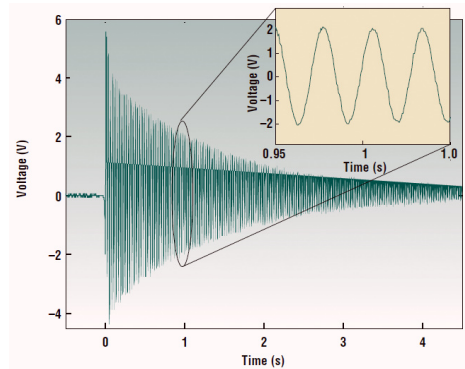

**Figure 8:** (left) A piezo electric material is attached to a cantilever so that it generates a voltage every time the cantilever oscillates to its mean position after it springs back. (right) The voltage response of the piezo electric material during the 'ring down' of cantilever. Reproduced from Ref. 24.

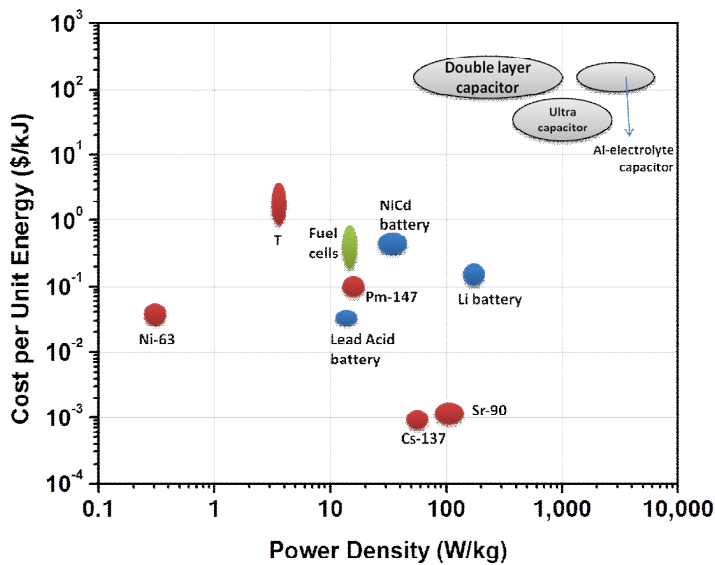

**Figure 9:** A plot of cost per unit energy as a function of power density for capacitors (gray), chemical batteries (blue), fuel cells (green) and atomic batteries made of various radioisotopes (red). Data collected from Refs. 26-31 and a market survey.

It is striking that Cs-137 and Sr-90 batteries stand out in terms of power density and cost; in fact, they top the charts of production among nuclear batteries. Many recent reports agree with this, atleast qualitatively.[14d] Some materials, like Tritium, have a wider range of cost depending on the production method. Sr-90 and Cs-137 are abundantly found in, or extracted from, the waste of nuclear fission, while Sr-90 is also naturally found in amounts of 300mg/kg of the earth's crust.[27] Apart from the cost, power and energy of the battery, the choice of material also depends on the type of radiation (alpha, beta, gamma) and the energy (or penetration capacity) of the radiation. We strongly believe that non-thermal converters using Sr-90 and Cs-137 will drive a major part of the market soon. We also feel that the regulations on the handling of these materials, and the trouble to overcome these to setup a processing facility, are obstacles that prevent these systems from competing (in market volumes) with mainstream chemical batteries.

## FUTURE WORK

Direct conversion nuclear batteries are identical to solar cells in their operation, except that they use high energy particles from radiation instead of photons. We have come across research that describe improvements in direct conversion nuclear batteries that use beta decay (beta-cells) by improving the diode structure (e.g.: using III-V materials, using p-i-n junctions),[14c, 14d] including mechanisms and structures for light trapping and preventing reflections,[14a, 14b, 32] using 3D geometries for light trapping[25, 32] and building beta-cells in porous silicon.[18] These methods of improving beta-cells are very similar to the ways in which the improvement of solar cells occurred.[33] Solar cells went the nanowires route after these improvements were made,[34] although the solar cell community has yet to make full use of carbon nanotubes, as it has done with nanowires, despite some efforts.[35, 36] Likewise, going the route of nanowires and nanotubes would improve the performance of beta-cells as nanowires did to solar cells. Typically, nanowires are used in a vertical configuration, as shown in Figure 10. The main advantages of using nanowires in this configuration are: (i) reduced reflection of incident particles or photons due to tapered refractive index, (ii) improved carrier collection because of radial extraction throughout the length, (iii) possibility of band structure tuning which reduces losses due to phonons and heat, and (iv) confinement of carriers in 1D, which could yield better transport. Some issues include surface impurities that degrade transport, as well as fabrication issues (i.e. conformal coating of the subsequent layers).

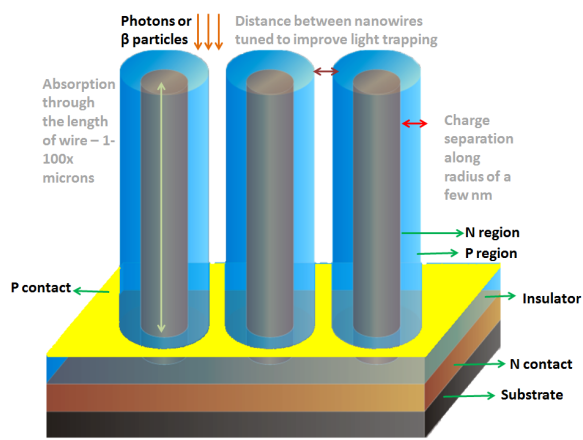

**Figure 10:** Configuration of a nanowires solar cell (or a beta-cell). Modified and adapted from Ref. 37.

Since the beta particles have far higher energy than photons, the beta-cell, unlike solar cells, must be designed so that the outer layer of the nanowires (generally p-type layer) is able to collect the thousands of excess electrons generated by each beta particle, and simultaneously be able to withstand large amounts of damage caused by the high energy impact. An alternative could be to fill in the space between the nanowires with radio luminescent substances like phosphor, protect the nanowires from beta particles, and use the light that phosphor emits upon impact to drive nanowire-based diodes. In this configuration, we can use low quality semiconductors that have a small diffusion length since the charge separation is along the radius of a few nm or more, or we could use a small bandgap material with very high electron mobility such as InSb.[15] In solar cells, similar configurations based on nanotubes have not yet been shown to outperform nanowires.

The aforementioned configuration is extremely hard to fabricate with nanotubes since they have very small diameter and no substantial distinction between the "core" and the surface of the nanotube, unlike that of p-n diode nanowires. It is also difficult to achieve the structural configuration of a nanowire array, which is tunable. Efforts in Refs. 35 and 36, however, show that it is possible to take this approach. Multi walled nanotubes (MWNTs), and specifically double walled nanotubes (DWNTs), can have distinct inner and outer walls that can be individually doped.[38, 39] The lengths of these can be up to a few hundreds of microns, while the spacing is mainly determined by the nature of the walls, with little influence by the growth conditions once the 'crowding effect' takes over and it is in the order of the diameter.[40] These smaller dimensions could be better for 'photonic crystal light trapping' of beta particles which have a smaller de-Broglie wavelength when compared to photons. Additionally, charge separation needs to occur within the small distance between the nanotube walls. Electron-hole pairs move as excitons because of their strong coupling in nanotubes. By fluorescence quenching, the exciton diffusion lengths can be reduced to the order of the distance between the walls.[41] The contacts to the two walls can be made with metals of different work functions, such that the band offsets allow only electrons to be collected from the n-type and holes from the p-type materials; thus, we could prevent shorting of the contacts even if they are in the physical vicinity of each other. We feel these improvements will give a quantum leap to direct and certain types of indirect conversion nuclear batteries.

## CONCLUSIONS

Direct conversion betavoltaic batteries and indirect conversion batteries can provide a few volts of open circuit voltage while the efficiency is less than 2%. Both suffer heavily from radiation-degradation if solid active material is used, which impedes electrical transport. RTGs can be loaded with a lot of material in one device, which is an advantage when we don't care about the weight of the device, hence the use of RTGs and radioisotope Sterling engines in space. Direct charge batteries do not suffer from material degradation and the battery's discharge follows the radioisotope decay. These can provide many kilovolts of potential at relatively higher efficiency. Their theoretical efficiency is high and they are able to operate in many extreme environmental conditions. Although MEMS based direct charge converters do not typically produce high voltages and the efficiency is relatively lower, because of the advancements in IC fabrication technology, the p-n junction based betavoltaic converter and the MEMS based piezoelectric battery show realistic promise for on-chip batteries that may incite a quantum leap for the IC industry. We compare the numbers but we do not claim one atomic battery to be generically better than another, since the choice depends on the requirements. Cost analysis shows that non-thermal converters are competitive in price ($/kJ), while those using Sr-90 and Cs-137 are far better in power density and $/kJ in comparison to chemical batteries. The hurdles in production and regulations have kept these from entering all domains of applications. Learning from the advancements in solar cells, use of nanomaterials could improve direct and indirect conversion batteries. Carbon nanotubes could prove more effective for beta-cells than they did for solar cells.

[1] H. G. J. Moseley, and J. Harling, The Attainment of High Potentials by the Use of Radium, Proc. R. Soc. (London) A, 88, 471 (1913).

[2] G Yakubov, Nuclear Batteries with Tritium And Promethium-147 Radioactive Sources, Doctoral Thesis, University of Illinois at Urbana-Champaign, 2010.

[3] L. Olsen, Betavoltaic energy conversion, Energy Convers. 13, 117 (1973).

[4] W. R. Corliss and D. G. Harvey, Radioisotopic Power Generation (Prentice-Hall, Englewood Cliffs, 1964).

[5] A. Lal, R. Duggirala, and H. Li, Pervasive power: a radioisotope-powered piezoelectric generator, IEEE Pervasive Computing, 4, 53 (2005).

[6] J. Lynch and K. Loh, A summary review of wireless sensors and sensor networks for structural health monitoring, Shock Vib. Dig. 38, 91 (2006).

[7] S. Shyamala, G. Udhayakumar, and A. Dash, Preparation 63 Ni electrodeposited special custom-made sources, BARC Newslett., no. 273, pp. 274–277, (2006).

[8] P.A. Karam and Ben P. Stein, Radioactivity (Infobase Publishing, 2009).

[9] T Christen and M Carlen, Theory of Ragone plots, Journal of Power Sources, 91, 2, 210-216 (2000).

[10] M. Romer, G. H. Miley, and R. J. Gimlin, Ragone Plot Comparison of Radioisotope Cells and the Direct Sodium Borohydride/Hydrogen Peroxide Fuel Cell With Chemical Batteries, IEEE Transactions on Energy Conversion 23 (1), 171 (2008).

[11] Y. V. Lazarenko, V. V. Gusev, and A. A. Pystovalov, Basic parameters of a radionuclide thermoelectric generator, Atomic Energy 64 (2), 131 (1988).

[12] D. J. Anderson, W. A. Wong, et al., An Overview and Status of NASA's Radioisotope Power Conversion Technology NRA, NASA/TM-2005-213980 (2005).

[13] Mark Wolverton, Stirling in Deep Space, Scientific American, 18 Feb 2008.

[14] Yuri A. Barbanel et al, Polymers, Phosphors, and Voltaics for Radioisotope Microbatteries (CRC Press, 2002). (a) Chapter 1 (b) Chapter 7 (c) Chapter 2 (d) Chapter 8

[15] David L. Pulfrey, Understanding Modern Transistors and Diodes (Cambridge University Press, 2010).

[16] V. Baranov, et al., Radioactive isotopes as energy sources in photovoltaic nuclear battery based on plasma-dusty structures, Isotopes, IzdAT (2000).

[17] W. Ehrenberg, et al., The Electron Voltaic Effect, Processing Royal Society 64, 424 (1951).

[18] Hang Guo et al, Betavoltaic microbatteries using porous silicon, Micro Electro Mechanical Systems, 2007. MEMS. IEEE 20th International Conference on, 867 – 870, Jan 2007.

[19] Peter Cabau et al, Micropower Betavoltaic Hybrid Sources, High Efficiency Energy Conversion, Energy Management, and Low Power Systems for Aerospace/Military Electronics Workshop, Redstone Arsenal, AL, Sep 2010.

[20] T. Wacharasindhu, et al, Radioisotope Microbattery Based on Liquid Semiconductor, Journal of Applied Physics Letters, 95, 014103 (2009).

[21] B. Liu, K. P. Chen, N. P. Kherany, et al., Betavoltaics using scandium tritide and contact potential difference, Applied Physics Letters 92, 083511 (2008).

[22] B. Gross, and P. V. Murphy, Currents from Gammas Make Detectors and Batteries, Nucleonics 19, 86 (1961).

[23] R Duggirala, A Lal and S Radhakrishnan, Chapter 4 - Radioisotope thin-film powered microsystems (Springer 2010).

[24] H. Li et al., Self-Reciprocating Radioisotope-Powered Cantilever, J. Applied Physics, 92, 2, 1122–1127 (2002).

[25] H Guo, and A La1, Nanopower Betavoltaic Microbatteries, The 12 1h International Conference an Solid Slate Sensors, Actuators and Microsystems, 36-39 (2003).

[26] L. J. Wittenberg, the cost of tritium production in a nuclear reactor, UWFDM-871, Nov 1991.

[27] R van Ginderdeuren, E van Limbergen, and W Spileers, 18 Years' experience with high dose rate strontium-90 brachytherapy of small to medium sized posterior uveal melanoma, Br J Ophthalmol, 89(10): 1306–1310 (2005).

[28] R. D. Rogers et al, Encapsulation method for maintaining biodecontamination activity, US Patent 7026146B2 (2006).

[29] D. J. Sims, Diffusion coefficients for uranium, cesium and strontium in unsaturated prairie soil, Journal of Radioanalytical and Nuclear Chemistry, Vol. 277, No.1, 143–147 (2008).

[30] Hoisington, J.E., Radioisotopes for heat-source applications, DPST-82-842 (1982).

[31] Baisden, A.C, ADVISOR-based model of a battery and an ultra-capacitor energy source for hybrid electric vehicles, Vehicular Technology, 53, 1, 199-205 (2004).

[32] W. Sun, et al, Harvesting Betavoltaic And Photovoltaic Energy With Three Dimensional Porous Silicon Diodes, Mat. Res. Soc. Symp. Proc. 836, 285-290 (2005).

[33] A Luque, Handbook of Photovoltaic Science and Engineering (John Wiley and Sons, 2011).

[34] E C Garnett et al, Nanowire Solar Cells, Annu. Rev. Mater. Res. 41:269–95 (2011).

[35] A Kongkanand, Single Wall Carbon Nanotube Scaffolds for Photoelectrochemical Solar Cells. Capture and Transport of Photogenerated Electrons, Nano Lett.., 7, 3, 676–680 (2007).

[36] J Wei, Double-Walled Carbon Nanotube Solar Cells, Nano Lett., 7, 8, 2317–2321 (2007).

[37] E C. Garnett et al, Silicon Nanowire Radial p−n Junction Solar Cells, J. Am. Chem. Soc., 130, 29, 9224–9225 (2008).

[38] F. Villalpando-Paez et al, Raman Spectroscopy Study of Isolated Double-Walled Carbon Nanotubes with Different Metallic and Semiconducting Configurations, Nano Lett., 8 11, 3879–3886 (2008).

[39] G Chen et al, Chemically Doped Double-Walled Carbon Nanotubes: Cylindrical Molecular Capacitors, Phys. Rev. Lett. 90, 257403 (2003).

[40] S Reich et al, Carbon Nanotubes: Basic Concepts and Physical Properties (Wiley-VCH, 2004).

[41] L Lüer et al, Size and mobility of excitons in (6, 5) carbon nanotubes, Nature Physics 5, 54 - 58 (2009).